\documentstyle[preprint,aps]{revtex}
\begin{document}
\title{Operator Ordering in Quantum Radiative Processes}
\author{J. L. Tomazelli}
\address{Departamento de F\'{\i}sica e Qu\'{\i}mica, 
%Faculdade de Engenharia,
Universidade Estadual Paulista, Campus da Guaratinguet\'a,
Av. Dr. Ariberto Pereira da Cunha 333,
12500-000 \\ Guaratinguet\'a, SP, Brazil.}
\author{L. C. Costa\footnote{{\it Corresponding Author}: Tel.: +55-11-3177-9023; Fax: +55-11-3177-9080\\
{\it E-mail addresses}: lccosta@ift.unesp.br (L.C. Costa), jeferson@camposc.feg.unesp.br (J.L. Tomazelli)}}
\address{Instituto de F\'{\i}sica Te\'{o}rica,
Universidade Estadual Paulista, \\
01405-900, S\~{a}o Paulo, Brazil.}
\maketitle
\begin{abstract}
In this work we reexamine quantum electrodynamics of atomic
electrons in the Coulomb gauge in the dipole approximation and
calculate the shift of atomic energy levels in the context of
Dalibard, Dupont-Roc and Cohen-Tannoudji (DDC) formalism by
considering the variation rates of physical observable. We then
analyze the physical interpretation of the ordering of operators in
the dipole approximation interaction Hamiltonian in terms of field
fluctuations and self-reaction of atomic electrons, discussing the
arbitrariness in the statistical functions in second order bound-state 
perturbation theory.\\
{\it PACS}: 05.10.Gg, 32.80.-t. \\
{\it Keywords}: density matrix, operator ordering, statistical functions.
\end{abstract}
%
%%%%%%%%%%%%%%%%%%%%%%%%%%%%%%%%%%%%%%%%%%%%%%%%%%%%%%%%%%%%%%%%%%%%%%%%%%%%%%%%%%%%%%%%%%%%%
%
\newpage
\section{Introduction}
In radiative processes, the ordering problem of atomic and field
operators in the interaction Hamiltonian of bound state QED has been
raised since the works by Senitzki\cite{SE73}, Ackerhalt {et
al}\cite{AC73} and others\cite{MI75}. Behind this discussion is the
physical interpretation of atomic radiative effects such as the
radiative line shifts in spontaneous emission. Alternative
approaches were proposed in order to elucidate important issues
concerning such problem. Among them are those based on the
complementarity between radiation reaction and vacuum fluctuation
effects, which provide a conceptual basis for the physical
interpretation of different radiative processes.

In the Dalibard, Dupont-Roc and Cohen-Tannoudji (DDC) formulation,
the ordering between the operators of the electromagnetic field,
considered as a reservoir (${\cal R}$), and a microscopic atomic
system (${\cal S}$) play a fundamental role in the identification of
the respective contributions due to the reservoir fluctuation (fr)
and the self-reaction (sr) \cite{CT82} - \cite{CT98}. They showed
that the symmetric ordering gives a true physical meaning to the
(fr) and (sr) rates.

In this article we study, in the context of the DDC construct, a more
general operator ordering and its physical significance to a given
observable variation rate, more specifically, an atomic energy
shift.

We use this analysis to establish a formal connection between DDC
approach and a closely related treatment, proposed by \cite{CT98},
which is based on the master equation formulation, where the
physical motivation relies on the classical theory of damping
harmonic oscillator \cite{LO73}. Finaly, we discuss the irrelevance 
of the ordering to an especific interaction Hamiltonian.
%
%%%%%%%%%%%%%%%%%%%%%%%%%%%%%%%%%%%%%%%%%%%%%%%%%%%%%%%%%%%%%%%%%%%%%%%%%%%%%%%%%%%%%%%%%%%%%
%
\section{The Effective Hamiltonian Formulation}
In the DDC construct the global Hamiltonian for a coupled system
${\cal S} + {\cal R}$ is, in the dipole approximation, given by
\begin{equation}
H=H_S+H_R+V,
\end{equation}
where $H_S$ is the Hamiltonian of the microscopic system ${\cal S}$,
$H_R$ the Hamiltonian of the reservoir ${\cal R}$ and $V$ the
interaction between ${\cal S}$ and ${\cal R}$, which we assume to be
of the form $V=-g RS$ ($g$ is the coupling constant and $R$ and $S$
are, respectively, Hermitian observable of ${\cal R}$ and ${\cal
S}$)\cite{CT82} - \cite{CT84}.

Following \cite{CT84} we set that the rate of variation for an
arbitrary Hermitian observable $G$ of ${\cal S}$ is given by the
Heisenberg equation of motion, and the contribution of the coupling
$V$ to this rate can be written as
\begin{equation}
\left( \frac{dG}{dt}\right)_{coupling} = - \frac{ig}{\hbar}  [R(t) S(t), G(t)] =
 g\lambda  N(t)R(t) + g(1-\lambda) R(t)N(t),
\end{equation}
where $N(t) = - (i/ \hbar) [S(t), G(t)]$ is an Hermitian observable
of the microscopic system and $\lambda$ an arbitrary real number. In
the above equation we have used the freedom in the ordering of
$R(t)$ and $N(t)$, since they commute.

In order to obtain the contributions of reservoir fluctuation (rf)
and self-reaction (sr) we perform the following replacement
\begin{equation}
X(t) = X^{\rm f} (t) + X^{\rm s} (t),
\end{equation}
($X = R, S, G$) where $R^f$ (resp. $S^f$ and  $G^f$) is the
solution, to order 0 in $g$, of the Heisenberg equation of motion
for $R$ (resp. $S$ and  $G$), corresponding to a free evolution
between $t_0$ and $t$, and $R^{\rm s} (t)$ (resp. $S^s$ and  $G^s$)
the solution to first order and higher in $g$. Then, substituting
(3) in (2) and retaining terms up to second order in $g$, we obtain
\begin{eqnarray}
\left( \frac{dG}{dt}\right)^{\rm rf} (t)&=& -\frac{ig}{\hbar}  \{
(1 - \lambda)R^{\rm f}(t)[S^{\rm f}(t), G^{\rm f}(t)] +
\lambda [S^{\rm f}(t), G^{\rm f}(t)]R^{\rm f}(t) \} - \nonumber \\
&-&\frac{g^2}{\hbar^2}  \int_{t_0}^t dt'[S^{\rm f}(t'),
[S^{\rm f}(t),G^{\rm f}(t)]]\times
\nonumber \\ & &\;\;\;\;\;\;\;\;\;\;\;\;\;\;\;\;\;\;\;\;\;\;\;\;\;\;\;\;\;\;\;
\times( (1-\lambda)R^{\rm f}(t')R^{\rm f}(t)) + \lambda R^{\rm f}(t)R^{\rm f}(t') ),\\
\left( \frac{dG}{dt}\right)^{\rm sr}(t) &=& -\frac{g^2}{\hbar^2}  \int_{t_0}^t dt'
[R^{\rm f}(t'),R^{\rm f}(t)] \times\nonumber \\
& &\;\;\;\;\;\;\;\;\;\;\;\;\times ((1-\lambda)S^{\rm f}(t')[S^{\rm f}(t),G^{\rm f}(t)]
+ \lambda[S^{\rm f}(t),G^{\rm f}(t)]S^{\rm f}(t')).
\end{eqnarray}
Since the rates (4) and (5) contain only free operators, their
average value in the reservoir state $\sigma_R$ gives\footnote{Note
that the term in the first line of (4) do not contribute to the
respective rate since it is linear in the absorption and emission
operators of the field.}
\begin{eqnarray}
\left< \left( \frac{dG}{dt} \right)^{\rm rf}(t) \right>^{(R)} &=&
-\frac{g'^2}{\hbar^2} \int_{t_0}^t dt' \; C^{(R)}(t, t', \lambda) \;
[S^{\rm f}(t'),[S^{\rm f}(t),G^{\rm f}(t)]], \\
\left< \left( \frac{dG}{dt} \right)^{\rm sr}(t) \right>^{(R)} &=&
-\frac{g'^2}{2\hbar^2} \int_{t_0}^t dt' \;
\chi^{(R)}(t, t') \times \nonumber \\
&\times& \{ (1-\lambda)S^{\rm f}(t')[S^{\rm f}(t),G^{\rm f}(t)]
+ \lambda[S^{\rm f}(t),G^{\rm f}(t)]S^{\rm f}(t') \},
\end{eqnarray}
where we have define $g'=\sqrt{2}g$ and
\begin{eqnarray}
C^{(R)}(t, t', \lambda) &=& \frac{1}{2} {\rm Tr_R} [ \sigma_{R} \{
\lambda R^{\rm f}(t)R^{\rm f}(t')  +  (1-\lambda)R^{\rm f}(t')R^{\rm f}(t) \} ], \\
\chi^{(R)}(t, t') &=& \frac{i}{\hbar} {\rm Tr_R}  \{ \sigma_{R}
[R^{\rm f}(t'),R^{\rm f}(t)] \} \theta(t-t').
\end{eqnarray}
The functions $C^{(R)}$ and $\chi^{(R)}$ are statistical functions
of the reservoir\cite{MA68} - \cite{HU56}. $C^{(R)}$ is a kind of
correlation function, describing the ``dynamics of fluctuations'' of
$R$ in the stationary state $\sigma_R(t_0)$; $\chi^{(R)}$ is the
linear susceptibility of the reservoir, determining the linear
response of the averaged observable $\langle R(t) \rangle$ when the
reservoir is acted upon by a perturbation\footnote{In (9) $\theta$
is the Heaviside function, $\theta(x) = 1$ if $x>0$, $\theta(x) = 0$
if $x<0$.}.

The above calculation has shown that the freedom in ordering (2)
just reflects in the correlation functions given by (8). This result
will be explored in section III, where we make a connection with the
master equation. In order to get a better understanding of the
aforementioned arbitrariness, we will consider the case of an atomic
energy shift.

In order to find the energy shifts corresponding to the (rf) and
(sr) rates we rewrite (6) and (7) in a convenient form, namely
\begin{eqnarray}
\left< \left( \frac{dG}{dt} \right)^{\rm rf}(t) \right>^{(R)} &=& \frac{i}{\hbar}
\langle [ (H_{eff}(t))^{\rm rf}, G(t) ] \rangle_{R} +  \\
&+& \left( \frac{-g'^2}{2 \hbar^2} \right) \langle [Y(t, \lambda),[S(t), G(t)]]
+ [S(t),[Y(t, \lambda), G(t)]] \rangle_{R}, \nonumber \\
\left< \left( \frac{dG}{dt} \right)^{\rm sr}(t) \right>^{(R)} &=& \frac{i}{\hbar}
\langle [ (H_{eff}(t))^{\rm sr}, G(t) ] \rangle_{\rm R} + \nonumber \\
&+& \left( \frac{- i g'^2}{4 \hbar^2} \right) \langle [{Z'}(t,\lambda)[S(t), G(t)]
+ {[}S(t), G(t){]}{Z''}(t,\lambda)- \nonumber \\ &-& S(t)[{Z''}(t,\lambda), G(t)] -
[{Z'}(t,\lambda), G(t)]S(t) \rangle_{R}
\end{eqnarray}
where
\begin{eqnarray}
(H_{eff}(t))^{\rm rf} &=& \frac{ig'}{2 \hbar}  [Y(t,\lambda), S(t)], \\
(H_{eff}(t))^{\rm sr} &=& \frac{-g'}{4}  [ Z'(t,\lambda) S(t) +
S(t) Z''(t,\lambda) ]
\end{eqnarray}
are second order corrections to the Hamiltonian part of ${\cal S}$
caused by its interaction with the reservoir and
\begin{eqnarray}
Y(t, \lambda) &=& \sum_{ab} q_{ab}(t) \langle a| S|b \rangle  \int_0^{\infty} d\tau \;
C^{(R)}(\tau, \lambda) e^{- i \omega_{ab} \tau}, \\
Z'(t,\lambda) &=& (1-\lambda) \sum_{ab} q_{ab}(t) \langle a| S|b \rangle
\int_{-\infty}^{\infty} d\tau \;
\chi^{(R)}(\tau) e^{- i \omega_{ab} \tau}, \\
Z''(t,\lambda) &=& \lambda \sum_{ab} q_{ab}(t) \langle a| S|b \rangle
\int_{-\infty}^{\infty} d\tau \;
\chi^{(R)}(\tau) e^{- i \omega_{ab} \tau}
\end{eqnarray}
with $q_{ab} \equiv |a \rangle \langle b|$, $\omega_{ab} = (E_a -
E_b)/\hbar$ and $\tau = t - t'$. Following the same point of view of
\cite{CT84}, expression (12) (resp. (13)) describes the part of the
evolution due to reservoir fluctuations (resp. due to self-reaction)
and which can be described by an effective Hamiltonian. The second
line of expression (10) (resp. (11)) describes the non-Hamiltonian
part of the evolution of $G$ caused by the reservoir fluctuation
(resp. self-reaction).
%
%%%%%%%%%%%%%%%%%%%%%%%%%%%%%%%%%%%%%%%%%%%%%%%%%%%%%%%%%%%%%%%%%%%%%%%%%%%%%%%%%%%%%%%%%%%%%
%
\subsection{The Energy Shifts: Hamiltonian Part}
Corrections (12) and (13) to the Hamiltonian $H_S$ affect ${\cal S}$
through a shifting in its energy eigenstates. Hence, considering a
state $|a\rangle$ (which is an eigenstate of $H_S$) we have the
following energy shifts
\begin{eqnarray}
(\delta E_a)^{\rm rf} &=& \langle a| (H_{eff}(t_0))^{\rm rf} |a \rangle, \\
(\delta E_a)^{\rm sr} &=& \langle a| (H_{eff}(t_0))^{\rm sr} |a \rangle.
\end{eqnarray}
Using expression (12), and noting that
\begin{equation}
Y(t_0) =  \int_0^{\infty}  C^{(R)} (\tau, \lambda) S^f (t_0 - \tau) d\tau,
\end{equation}
expression (17) for  $(\delta E_a)^{\rm rf}$ becomes
\begin{equation}
(\delta E_a)^{\rm rf}  =  - \frac{g'^2}{2}
\int_{- \infty}^{+ \infty} C^{(R)} (\tau, \lambda) \chi^{(S,a)} (\tau) d\tau,
\end{equation}
where we have introduced a new statistical function, the
susceptibility of the system obser\-vables
\begin{equation}
\chi^{(S,a)} (\tau) = \frac{i}{\hbar} \langle a| [S^f(t_0), S^f(t_0 - \tau) ] |a \rangle
\theta(\tau).
\end{equation}
From expression (13) for $(H_{eff})^{\rm sr}$, we can follow the same steps as those
from (17) to (20). As a result we obtain
\begin{equation}
(\delta E_a)^{\rm sr} =  - \frac{g'^2}{2}  \int_{- \infty}^{+ \infty}
\chi^{(R)} (\tau) C^{(S,a)} (\tau, \lambda) d\tau,
\end{equation}
where, again, we have introduced a new statistical function, the
``correlation'' for the system observable
\begin{equation}
C^{(S,a)} (\tau, \lambda) = \frac{1}{2} \langle a| \lambda S^f(t_0) S^f(t_0 - \tau)
 + (1-\lambda) S^f(t_0 - \tau) S^f(t_0)|a \rangle.
\end{equation}

For future convenience we write (20) and (22) in the frequency
space. Using the Parseval's theorem we have
\begin{equation}
{(\delta E_a)}^{\rm rf}  =  - \frac{g'^2}{2} \int_{- \infty}^{+ \infty} 
d\omega \; C^{(R)}(\omega, \lambda) \chi^{(S,a)} (\omega) ,
\end{equation}
\begin{equation}
(\delta E_a)^{\rm sr} =  - \frac{g'^2}{2}  \int_{- \infty}^{+ \infty}
d\omega \; \chi^{(R)} (\omega) C^{(S,a)} (\omega, \lambda),
\end{equation}
where we have used the parity properties of $C$ and $\chi$ \cite{CT84}.

Formulas (24) and (25) give us the energy shifts which, {\it a
priori}, depends on $\lambda$ through the ``correlation functions'',
expressions (8) and (23). DDC argued that a true physical meaning is
obtained by choosing $\lambda=1/2$ since then both variation rates,
expressions (6) and (7), become hermitian quantities. As a consequence 
the energy shifts (24) and (25) will correspond to the (rf) and (sr) 
effects. It will be shown in section IV that the above assertion becomes 
meaningless in the case where the interaction Hamiltonian is of 
the form  $V \sim {\bf p} \cdot {\bf A}$.
%
%%%%%%%%%%%%%%%%%%%%%%%%%%%%%%%%%%%%%%%%%%%%%%%%%%%%%%%%%%%%%%%%%%%%%%%%%%%%%%%%%%%%%%%%%%%%%%
%
\section{The Master Equation Approach}
In this section, we use the previous results to establish a formal 
connection with the approach employed in \cite{CT98}, where the 
physical interpretation of the energy shifts in term {\rm (rf)} and 
{\rm (sr)} are obtained without any reference to operator ordering. 
In fact, the same energy shifts given by (24) and (25) can also be 
obtained using a matrix approach based on the evolution equation for 
the density operator of the global system ${\cal S} + {\cal R}$ in the 
interaction picture with respect to $H_S + H_R$. Hence, following 
\cite{CT98}, the energy shift for a state $|a\rangle$ of ${\cal S}$ 
caused by its interaction with ${\cal R}$ through $V$ is given by
\begin{equation}
\Delta_a=\frac{1}{\hbar}{\cal P}\sum_{\mu, \nu}p_{\mu}\sum_b
\frac{|\langle\nu,b|V|\mu,a\rangle|^2}{E_{\mu}+E_a-E_{\nu}-E_b}
\end{equation}
where $p_{\mu}$ is a distribution of probability corresponding to
the reservoir average in the stationary state $\sigma_R$ and
$|\mu\rangle$, $|\nu\rangle$ are eigenstates of $H_R$ with eigenvalue
$E_{\mu}$, $E_{\nu}$. In (26) ${\cal P}$ denotes the principal value.

It is directely to see that the matrix element $\langle\mu,a|V|\nu,b\rangle$ 
in (26) can be factorized in two parts, one relative to ${\cal S}$ and another 
relative to ${\cal R}$,
\begin{equation}
\Delta_a=\frac{g'^2}{2 \hbar^2}\sum_{\mu, \nu}p_{\mu}| \langle \mu|R|\nu \rangle |^2\left[
\sum_b| \langle a|S|b \rangle |^2{\cal P}\frac{1}{\omega_{\mu\nu}+\omega_{ab}}\right].
\end{equation}
In this way, since we know the functional structure of $C(\omega)$ and $\chi(\omega)$ 
\cite{CT98}, namely
\begin{equation}
C^{(R)}(\omega)=\sum_{\mu, \nu}p_{\mu}\pi |\langle \mu|R|\nu \rangle|^2
[\delta(\omega+\omega_{\mu\nu})+\delta(\omega-\omega_{\mu\nu})],
\end{equation}
\begin{equation}
\chi^{(R)}(\omega)={\chi'}^{(R)}(\omega)+i {\chi''}^{(R)}(\omega)\,\,,
\end{equation}
\begin{eqnarray}
{\chi'}^{(R)}(\omega)=-\frac{1}{\hbar}\sum_{\mu, \nu}p_{\mu} |\langle \mu|R|\nu \rangle|^2
\left[{\cal P}\frac{1}{\omega_{\mu\nu}+\omega}+
{\cal P}\frac{1}{\omega_{\mu\nu}-\omega}\right] , \\
{\chi''}^{(R)}(\omega)=\frac{\pi}{\hbar}\sum_{\mu, \nu}p_{\mu}
|\langle \mu|R|\nu \rangle|^2 \left[\delta(\omega_{\mu\nu}+\omega)-\delta(\omega_{\mu\nu}-
\omega)\right],
\end{eqnarray}
and analogous expressions for ${\cal S}$ (where only $p_a=1$ is nonzero),
we can make a mathematical trick and rewrite the fraction $1/(\omega_{\mu\nu}+\omega_{ab})$ as
\begin{eqnarray}
{\cal P}\frac{1}{\omega_{\mu\nu}+\omega_{ab}} &=& \frac{1}{2}\int d\omega \times \nonumber \\
&{\times}& \left\{ \left({\cal P}\frac{1}{\omega_{\mu\nu}+\omega}+
{\cal P}\frac{1}{\omega_{\mu\nu}-\omega}\right)
[\lambda\delta(\omega+\omega_{ab})+(1-\lambda)\delta(\omega-\omega_{ab})]
+ \right. \nonumber \\
&+& \left. \left({\cal P}\frac{1}{\omega_{ab}+\omega}+
{\cal P}\frac{1}{\omega_{ab}-\omega}\right)
[\lambda \delta(\omega+\omega_{\mu\nu})+(1-\lambda) \delta(\omega-\omega_{\mu\nu})] \right\}.
\end{eqnarray}
The presence of the $\lambda$ parameter in the correlation functions of equations 
(24) and (25) suggests the above construction. This point is crucial because it 
clarifies the motivation behind the use of (32) in the present context.

Now, substituting (32) into (27) we obtain: $\Delta_a = \Delta_a^{\rm rf}
+ \Delta_a^{\rm sr}$, where
\begin{eqnarray}
\hbar\Delta_a^{\rm rf} &=& -\frac{{g'}^2}{2} \int_{-\infty}^{\infty}
\frac{d\omega}{(2 \pi)} \; C^{(R)} (\omega, \lambda) \chi^{(S,a)}(\omega) , \\
\hbar\Delta_a^{\rm sr} &=& -\frac{{g'}^2}{2}\int_{-\infty}^{\infty} 
\frac{d\omega}{(2 \pi)} \; \chi^{(R)}(\omega) C^{(S,a)}(\omega, \lambda).
\end{eqnarray}

Chosing $\lambda = 1/2$ in the above expression the original formulation given 
in \cite{CT98} is recovered and the physical meaning of (27) becomes simple and 
clear in terms of (rf) and (sr) effects. This are in complete agreement with 
the results obtained in the last section, expressions (24) and (25). However, in 
this approach \cite{CT98} the physical interpretation are borrowed from the classical 
theory of damping harmonic oscillator, without any reference to the operator ordering 
in the correlation functions. In fact, for a class of interaction Hamiltonian of the 
form $V \sim {\bf p} \cdot {\bf A}$ the choice of $\lambda$ in (33) and (34) (or 
equivalently (24) and (25)) becomes irrelevant as will be shown in the next section. 
%
%%%%%%%%%%%%%%%%%%%%%%%%%%%%%%%%%%%%%%%%%%%%%%%%%%%%%%%%%%%%%%%%%%%%%%%%%%%%%%%%%%%%%%%%%%%%%
%
\section{The Dipole Interaction}
In order to show the independence of (33) (or (34)) with respect to $\lambda$ 
we consider here the same case treated in \cite{CT98} where the interaction 
Hamiltonian reduces to the expression
\begin{equation}
V = -\sum_i \left(\frac{e}{m}p_i\right)A_i({\bf r=0})\,, 
\end{equation}
where $i=x,y,z$. Comparing the above expression with the definition of $V$ 
($V=-g \sum_i R_i S_i$) we see that $g=1$, $R_i = A_i(\bf r=0)$ (where ${\bf r=0}$ 
reflects the long wavelenght approximation we are assuming) and $S_i = (e p_i/ m)$. 

In this context, the final expression for the $x$-componet of the statistical 
functions are
\begin{eqnarray}
\hat{C}_R^{xx}(\omega)&=&\frac{1}{3\pi\varepsilon_0c^3}
\int_0^{\omega_M}d\omega' \hbar\omega'( \langle n(\omega')\rangle + 1/2 )
[ \lambda \delta(\omega'+\omega)+ (1 - \lambda) \delta(\omega'-\omega) ], \\
\hat{\chi}_R^{'xx}(\omega)&=&\frac{1}{6\pi^2\varepsilon_0c^3}
\int_0^{\omega_M}d\omega'\omega'\left[{\cal P}\frac{1}{\omega'+\omega}+{\cal P}
\frac{1}{\omega'-\omega}\right], \\
\hat{\chi}_R^{''xx}(\omega)&=&\frac{-1}{6\pi\varepsilon_0c^3}
\int_0^{\omega_M}d\omega'\omega'[\delta(\omega'+\omega) - \delta(\omega'-\omega)],
\end{eqnarray}
and 
\begin{eqnarray}
\hat{C}_{Aa}^{xx}(\omega)&=&\sum_b\frac{e^2}{m^2}|\langle a|p_x|b\rangle |^2\pi
[\lambda \delta( \omega_{ab}+\omega)+ (1 - \lambda) \delta(\omega_{ab}-\omega)], \\
\hat{\chi}_{Aa}'^{xx}(\omega)&=&\sum_b\frac{-e^2}{\hbar m^2}|\langle a|p_x|b\rangle |^2
\left[ {\cal P}\frac{1}{\omega_{ab}+\omega}+{\cal P}\frac{1}{\omega_{ab}-\omega}\right], \\
\hat{\chi}_{Aa}''^{ xx}(\omega)&=&\sum_b\frac{e^2}{\hbar m^2}|\langle a|p_x|b\rangle |^2
\pi[\delta(\omega_{ab}+\omega)-\delta(\omega_{ab}-\omega)],
\end{eqnarray}
where $|a\rangle$ denotes a given eigenstate of $H_S$. The above expression are obtained 
following reference \cite{CT98} and tanking into account identity (32). 

In order to obtain the energy shift $\Delta_a^{\rm rf}$ we substitute (36) and (40) 
in (33). Remembering that $g'= \sqrt{2} g = \sqrt{2}$, we obtain
\begin{eqnarray}
\hbar \Delta_a^{\rm rf}&=& - \int_{\infty}^{\infty} \frac{d \omega}{2 \pi} 
\left\{ -\frac{1}{\hbar} \sum_b \frac{e^2}{m^2} | \langle a | p_x | b \rangle |^2 
\left[ {\cal P}\frac{1}{\omega_{ab}+\omega}+{\cal P}\frac{1}{\omega_{ab}-\omega}\right] \right\} 
\times \nonumber \\
&\times& \left\{ \frac{1}{3\pi\varepsilon_0c^3} \int_0^{\omega_M} d\omega' \hbar\omega' 
( \langle n(\omega')\rangle + 1/2 ) 
[ \lambda \delta(\omega'+\omega)+ (1 - \lambda) \delta(\omega'-\omega) ]  \right\}.
\end{eqnarray}
Performing the integral over $\omega$, it is straightforward to see that the contribuction 
from $\lambda$ and that from $- \lambda$ will cancel each other remaning just the contribution 
from the independent term, namely,
\begin{eqnarray}
\hbar \Delta_a^{\rm rf} = \frac{e^2}{6 \pi \varepsilon_0 m^2 c^3} 
\sum_b | \langle a | {\bf p} | b \rangle |^2 
\int_0^{\omega_M} d\omega' \omega' ( \langle n(\omega')\rangle + 1/2 ) 
\left[ {\cal P}\frac{1}{\omega_{ab}+\omega'}+{\cal P}\frac{1}{\omega_{ab}-\omega'}\right].
\end{eqnarray}
where the analogous contribution coming from the $y$ and $z$ components have already 
been added. As we have seen, the above construct clearly shows the independence of (33) 
with respect to $\lambda$. In the original formulation given in \cite{CT98} the above 
expression may indeed be divided in two parts: one proportional to the factor $1/2$ and 
associated to the Lamb-Retherford energy shift and another proportional to photon nunber 
and associated to the AC Stark effect. 

It must be also noted that the choice $\lambda=1/2$ made in the operatorial approach may be 
seen as a fixing parameter which ultimately gives an hermitian character to the contributions 
(6) and (7) allowing a fictious physical meaninging for (24) and (25) in terms of (rf) and (sr) 
effects. However, as remarked by Milonni \cite{MI94}, a true physical meaning is completely 
arbitrary since the observable variation rates are, in fact, unaccessible for the experiment.
%
%%%%%%%%%%%%%%%%%%%%%%%%%%%%%%%%%%%%%%%%%%%%%%%%%%%%%%%%%%%%%%%%%%%%%%%%%%%%%%%%%%%%%%%%%%%%%
%
\section{Concluding Remarks}
In this work we have applied to the original formulation of DDC construct 
a more general ordering between the atomic and electromagnetic field 
(reservoir) operators and calculated the energy shift due to the effective 
Hamiltonian part. The result showed that the freedom in ordering expression 
(2) reflects in the energy shifts (24) and (25) through the $\lambda$'s 
appearance in the correlation functions. 

We have also established a formal connection between the Effective
Hamiltonian approach and that based on the master equation theory.
This connection was guided by the results concerning the general
ordering (2) and consequently made explicit the same kind of
arbitrariness in the physical interpretation of the related 
variation rates (i.e., the energy shifts (33) and (34)) of section III.
In the sequence, it was also discussed that, for an interaction 
Hamiltonian of the form 
\begin{equation}
V = -\sum_i \left(\frac{e}{m}p_i\right)A_i({\bf r=0})\,, 
\end{equation}
the ordering (2) (or (4) and (5)) becomes meaningless and its role 
on the final result (expression (43)) turn out to be irrelevant. As 
a consequence the Lamb-Retherford shift and the AC Stark effect may 
be obtained without any particular ordering \cite{LU00}.

Another interesting result relies on the fact that our procedure 
still permit us to fix {\it a posteriori} a suitable ordering 
which keeps its (rf) and (sr) interpretation, as can be seen by 
looking directly to expressions (33) and (34). However, since 
the Hermicity of (33) and (34) holds only for $\lambda = 1/2$, 
such interpretation becomes artificial.

Once we get a better understanding on the arbitrariness in the
operator ordering in DDC construct, we expect to find a direct
connection with the works by Senitzki, Ackerhalt and others. The
main idea is to construct a similar structure in Fock space and
analyze its connection with all possible physical interpretations.

Another interesting application of the present formalism is a
possible generalization of the operator ordering in the spirit of
q-deformed operator algebras \cite{BO99}, subject of a forthcoming 
work.
%%%%%%%%%%%%%%%%%%%%%%%%%%%%%%%%%%%%%%%%%%%%%%%%%%%%%%%%%%%%%%%%%%%%%%%%%%%%%%%%%%%%%%%%%%%%
%
\section{Acknowledgements}
JLT thanks CNPq for partial financial support and the IFT/UNESP for
the hospitality. LCC is grateful to FAPESP for the financial
support.
%
%%%%%%%%%%%%%%%%%%%%%%%%%%%%%%%%%%%%%%%%%%%%%%%%%%%%%%%%%%%%%%%%%%%%%%%%%%%%%%%%%%%%%%%%%%%%%%
%

%
%%%%%%%%%%%%%%%%%%%%%%%%%%%%%%%%%%%%%%%%%%%%%%%%%%%%%%%%%%%%%%%%%%%%%%%%%%%%%%%%%%%%%%%%%%%%
%
\end{document}